\documentclass[aps,pra,reprint,superscriptaddress,showpacs,amsmath,amssymb]{revtex4-2}

\usepackage{graphicx}  
\usepackage{dcolumn}   
\usepackage{bm}        
\usepackage{hyperref}  
\usepackage{subcaption}
\usepackage[justification=raggedright,singlelinecheck=false]{caption}
\usepackage{booktabs}
\usepackage{siunitx}
\hypersetup{colorlinks=true, linkcolor=blue, citecolor=blue, urlcolor=blue}

\begin{document}

\title{Discrete Cavity Dynamics in Free-Space Brillouin Laser}

\author{Jiabao Peng}
\thanks{These authors contributed equally to this work.} 
\affiliation{Helmholtz-Institut Jena, Fröbelstieg 3, 07743 Jena, Germany}
\affiliation{Abbe Center of Photonics, Friedrich Schiller University Jena, Albert-Einstein-Straße 6, 07745 Jena, Germany}

\author{Longjie Zhang}
\thanks{These authors contributed equally to this work.}
\affiliation{Center for Advanced Laser Technology, Hebei University of Technology, Tianjin 300401, China}
\affiliation{Hebei Key Laboratory of Advanced Laser Technology and Equipment, Tianjin 300401, China}
\affiliation{Collaborative Innovation Center for Diamond Laser Technology and Applications, Tianjin 300401, China}
  
\author{Zhenxu Bai}
\email{baizhenxu@hotmail.com }
\affiliation{Center for Advanced Laser Technology, Hebei University of Technology, Tianjin 300401, China}
\affiliation{Hebei Key Laboratory of Advanced Laser Technology and Equipment, Tianjin 300401, China}
\affiliation{Collaborative Innovation Center for Diamond Laser Technology and Applications, Tianjin 300401, China}

\author{Stephan Fritzsche}
\email{s.fritzsche@gsi.de}
\affiliation{Helmholtz-Institut Jena, Fröbelstieg 3, 07743 Jena, Germany}
\affiliation{Theoretisch-Physikalisches Institut, Friedrich-Schiller-Universitat Jena, D-07743 Jena, Germany}

\author{Zhiwei Lu}
\email{zhiweilv@hebut.edu.cn}
\affiliation{Center for Advanced Laser Technology, Hebei University of Technology, Tianjin 300401, China}
\affiliation{Hebei Key Laboratory of Advanced Laser Technology and Equipment, Tianjin 300401, China}
\affiliation{Collaborative Innovation Center for Diamond Laser Technology and Applications, Tianjin 300401, China}

\begin{abstract}
Highly coherent lasers are central to modern photonics. 
To date, high-coherence operation has been achieved predominantly in microcavity and fiber-based platforms.   
More recently, free-space Brillouin-laser experiments have revealed unusually strong noise suppression whose physical origin cannot 
be explained by conventional continuous-medium models developed for those platforms.
In conventional continuous-medium models, the optical and acoustic fields are assumed to remain continuously coupled throughout the 
cavity evolution, whereas in free-space implementations the coupling is confined to the nonlinear medium and interrupted by passive 
propagation over the rest of the round trip.
To describe this interaction–propagation separation, we develop a discrete-cavity model in which the short Brillouin interaction 
inside the gain medium and the subsequent free-space propagation are treated as two separate stages of the round-trip evolution.
This separation introduces a temporal asymmetry between optical storage and acoustic relaxation, which effectively enhances acoustic damping at the cavity level and strongly reduces pump-noise transfer to the Stokes field.
If the cavity round-trip time is much longer than the interaction time in the nonlinear medium, the noise-suppression ratio scales with the ratio of the total cavity length to the nonlinear-medium length.
Our discrete-cavity model further provides quantitative predictions for the lasing threshold, output power, phase-noise transfer, 
and fundamental linewidth, in good agreement with experiment. These results identify the discrete interaction-propagation
structure as the physical origin of the unusually strong noise suppression in free-space Brillouin lasers systems.
\end{abstract}
\maketitle

\section{Introduction}
Highly coherent lasers are essential for a wide range of precision measurement applications,
including optical atomic clocks \cite{Loh_2020,bloom2014optical}, quantum information processing \cite{haffner2008quantum,su2016quantum}, gravitational-wave 
detection \cite{lancedigo2015adv,abbott2016observation}, and high-resolution spectroscopy \cite{rafac2000sub}.
Conventional routes to narrow-linewidth and low-noise laser operation include frequency stabilization with 
high-$Q$ cavities \cite{matei20171}, distributed-feedback architectures \cite{jin2021hertz}, non-planar 
ring oscillators \cite{kane1985monolithic}, and stimulated Brillouin scattering (SBS) \cite{pant2011chip}. 
In practice, many high-coherence laser systems are realized in waveguide-based or other monolithic platforms, such as semiconductor 
devices, optical fibers, and on-chip integrated structures, where the optical field remains continuously embedded in the underlying 
medium.
Among the various narrow-linewidth laser technologies, Brillouin lasers have emerged as a particularly important platform
\cite{tomes2009photonic,lee2012chemically,kabakova2013narrow,gundavarapu2019sub,mao2025narrow,
chauhan2021visible}. 
The coherence enhancement in Brillouin lasers arises because the acoustic damping is much stronger than the optical damping.
As a result, fast pump phase fluctuations cannot efficiently drive the acoustic mode, and pump-noise transfer to the Stokes 
field is strongly suppressed.
This mechanism leads to substantial linewidth narrowing and reduced noise. 
Early Brillouin lasers were realized mainly in optical fibers 
\cite{hill1976cw,stokes1982all,smith1991narrow}, where long interaction lengths and strong confinement 
enable efficient SBS at low threshold.
More recently, microcavity Brillouin lasers \cite{perin2025hz,chen2024stabilized,hu2014low,loh2019ultra,tao2022single} 
have since demonstrated exceptionally narrow linewidths and strong noise suppression within spatially continuous 
cavity-media configurations \cite{debut2000linewidth}.

Whereas continuous-medium Brillouin lasers have achieved narrow linewidths and strong noise suppression in monolithic platforms,
a recent free-space Brillouin-laser experiment \cite{zhang2025frequency} reported remarkably low noise levels and strong noise suppression, pointing to the 
potential of this platform for high-power, ultra-low-noise operation.
In particular, the observed noise-suppression ratios substantially 
exceed the predictions of conventional continuous-medium theories that have successfully described 
fiber and microresonator Brillouin lasers \cite{debut2000linewidth,loh2015noise}. 
In these conventional models, the optical and acoustic fields are treated as co-propagating and continuously coupled within the same 
nonlinear gain medium throughout the cavity evolution. 
This framework successfully accounts for linewidth 
narrowing, noise suppression, and noise transfer in guided Brillouin-laser platforms. 
In a free-space 
Brillouin laser, however, the optical and acoustic fields no longer co-propagate and remain continuously coupled throughout the cavity 
round trip: the optical field interacts with the acoustic field only during the short passage through the gain medium, whereas a 
substantial fraction of each round trip consists of passive free-space propagation.
During the passive free-space propagation stage, the optical field continues to propagate, while the acoustic field remains confined 
to the gain medium and relaxes independently.
Conventional continuous-medium models therefore fail to reproduce 
the experimentally observed degree of noise suppression in free-space systems, 
suggesting that key physical mechanisms are not captured in existing theoretical descriptions.

To describe free-space Brillouin lasers theoretically, we develop a discrete-cavity model.
In Sec.~II, we first review the conventional continuous-medium description of the three-wave coupled equations for Brillouin interactions, 
in which the evolution of the optical
and acoustic fields is governed by coupled differential equations. 
In Sec.~III, we introduce the discrete-cavity formulation, where the continuous time derivatives in the 
conventional model are replaced by a round-trip iterative map. This discrete-cavity formulation naturally 
incorporates the free-space propagation between successive interactions in the gain medium. 
In Sec.~IV, we solve the discrete iterative model by determining its fixed points and derive the 
corresponding threshold condition. 
In Sec.~V, we then analyze the noise properties of the system by applying 
a discrete Fourier transform to the linearized dynamics to derive the noise-transfer relations and evaluate both the relative intensity 
noise and the fundamental
linewidth.
Finally, Sec.~VI summarizes the main results and discusses their implications.

\section{Continuous-Medium Model for Brillouin Lasers}
\label{second chapter}
\begin{figure*}[t]
\centering

\begin{subfigure}{0.45\textwidth}
\centering
\includegraphics[width=\linewidth]{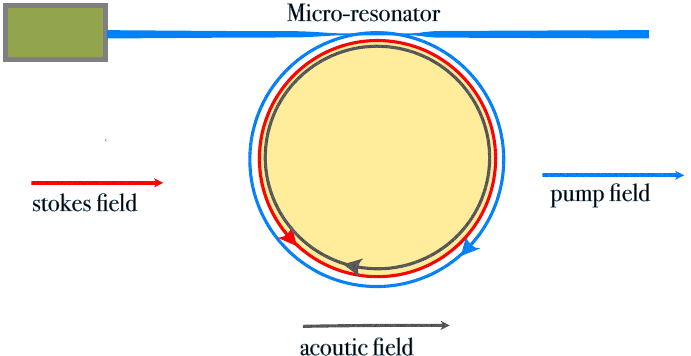}
\end{subfigure}
\hfill
\begin{subfigure}{0.45\textwidth}
\centering
\includegraphics[width=\linewidth]{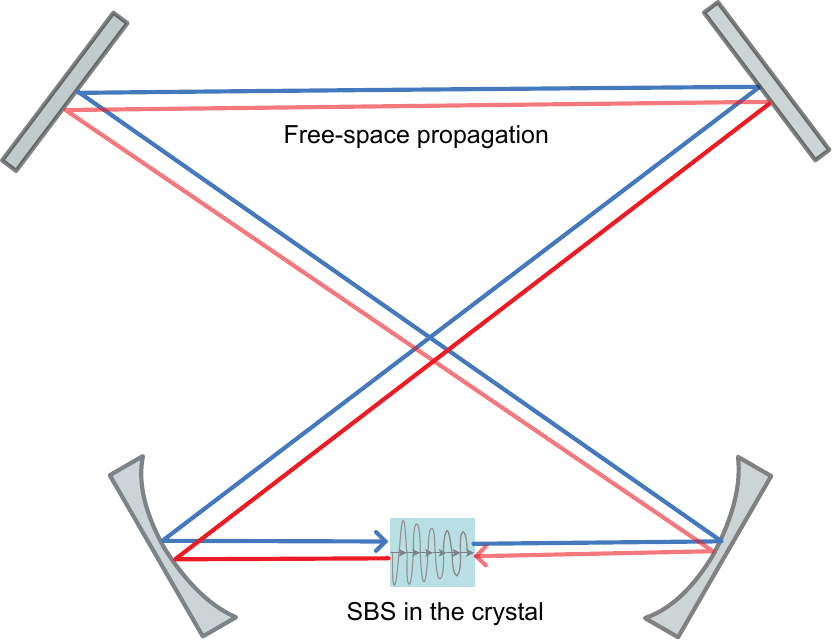}
\end{subfigure}
\caption{
Comparison of the conventional continuous-medium scheme and the free-space Brillouin laser configuration.
(a) Microresonator Brillouin laser with continuous opto-acoustic interaction in a co-localized nonlinear medium.
(b) Free-space Brillouin laser with localized interaction in the gain crystal and passive propagation over the remainder of the cavity round trip.
}
\label{fig:model_compare}
\end{figure*}
We begin from the electromagnetic wave equation in a dielectric medium,
\begin{equation}
\nabla^2 \mathbf{E}
-\frac{n^2}{c^2}\frac{\partial^2 \mathbf{E}}{\partial t^2}
=
\frac{1}{\epsilon_0 c^2}\frac{\partial^2 \mathbf{P}}{\partial t^2},
\label{eq:wave_equation}
\end{equation}
where \(\mathbf{E}\) is the electric field, \(n\) is the refractive index
of the medium, \(c\) is the speed of light in vacuum, and
\(\epsilon_0\) is the vacuum permittivity. The term \(\mathbf{P}\)
denotes the electrostrictively induced nonlinear polarization.
For the SBS process, 
the optical field is decomposed into forward-propagating pump and
backward-propagating Stokes components with slowly varying envelopes,
\begin{equation}
\begin{aligned}
E_p(\mathbf{r},t)
&= \frac{1}{2} \psi_{p}(x,y) A_p(z,t)e^{i(k_p z-\omega_p t)}+\mathrm{c.c.} \\
E_s(\mathbf{r},t)
&= \frac{1}{2} \psi_{s}(x,y) A_s(z,t)e^{i(-k_s z-\omega_s t)}+\mathrm{c.c.}
\end{aligned}
\label{eq:optical_ansatz}
\end{equation}
The Brillouin interaction originates from the
electrostrictive modulation of the refractive index, which gives rise to the
nonlinear polarization
\begin{equation}
P(\mathbf{r},t)=\epsilon_0\rho_0^{-1}\gamma_e\,\psi_{\rho}(x,y)\tilde{\rho}(z,t)\,E(\mathbf{r},t)
\label{eq:electrostrictive_polarization}
\end{equation}
The acoustic density fluctuation is represented by a slowly varying envelope,
\begin{equation}
\tilde{\rho}(z,t)=\rho_0+\left[\rho(z,t)e^{i(qz-\Omega t)}+\mathrm{c.c.}\right]
\label{eq:acoustic_ansatz}
\end{equation}
where \(\rho_0\) denotes the equilibrium mass density of the medium,
\(\rho(z,t)\) represents the slowly varying complex amplitude of the acoustic
density wave, and \(\gamma_e\) is the electrostrictive constant.

Substituting Eqs.~(\ref{eq:optical_ansatz}) and (\ref{eq:electrostrictive_polarization})
into Eq.~(\ref{eq:wave_equation}), and applying the slowly varying envelope
approximation, we retain only the resonant driving terms that satisfy the
energy and momentum conservation conditions of the SBS process. These
conditions require \(\omega_p=\omega_s+\Omega\) and \(k_p=k_s+q\), where
\(\Omega\) and \(q\) denote the angular frequency and wave vector of the
acoustic wave, respectively. Under these conditions, the coupled dynamics
reduce to the standard three-wave interaction equations for the pump, Stokes,
and acoustic fields.
In this process, the interference between the pump and Stokes fields produces a beat 
note that drives a coherent acoustic wave through electrostriction. 
The resulting 
density modulation, in turn, changes the refractive index of the medium and 
scatters pump photons into the Stokes mode. The SBS process can therefore be 
viewed as a closed three-wave coupling, in which optical energy is transferred 
from the pump field to the Stokes field through the mediation of the acoustic wave.
So the three coupled wave equations have the form of 
\begin{equation}
\setlength{\jot}{12pt}
\begin{aligned}
\dfrac{\partial A_{p}}{\partial t}
&=
-\dfrac{\gamma_{p}}{2} A_{p}
+
\dfrac{i\omega_{p}\gamma_{e}}{4 n^2 \rho_{0}}\,
\Lambda_{p}\,\rho\, A_{s}
+
\sqrt{\kappa_{ex}}S_{\mathrm{in}}
\\
\dfrac{\partial A_{s}}{\partial t}
&=
-\dfrac{\gamma_{s}}{2} A_{s}
+
\dfrac{i\omega_{s}\gamma_{e}}{4 n^2 \rho_{0}}\,
\Lambda_{s}\,\rho^{*}\, A_{p}
\\
\dfrac{\partial \rho}{\partial t}
&=
-\dfrac{\Gamma_{B}}{2}\,\rho
+
\dfrac{i\epsilon_{0}\gamma_{e}^{2}q^2}{4\Omega}\,
\Lambda_{\rho}\, A_{p}A_{s}^{*}
\end{aligned}
\label{three_coupled_wave_equations}
\end{equation}
In Eqs.~(\ref{three_coupled_wave_equations}),$\gamma_{p}$ and $\gamma_{s}$ denote the total decay
rates of the pump and Stokes cavity modes, respectively. These rates include both intrinsic
cavity losses and external coupling losses and therefore determine the photon lifetimes of the
corresponding intracavity optical fields. 
The parameter $\Gamma_{B}$ represents the damping rate of the acoustic wave, which describes the
relaxation of the acoustic field due to acoustic dissipation in the medium. 
The term $\sqrt{\kappa_{ex}}S_{\mathrm{in}}$ describes the injection of the external pump field into the cavity, 
where $S_{\mathrm{in}}$ denotes the input pump field and $\kappa_{ex}$ is the external coupling rate between the 
pump source and the cavity mode. This source term follows the standard input--output formulation of temporal 
coupled-mode theory, in which the external driving field enters the cavity equation through the factor 
$\sqrt{\kappa_{ex}}$ \cite{haus1983waves}. In the normalization adopted here, the intracavity field 
amplitudes $A_p$ and $A_s$ have units of $\sqrt{\mathrm{energy}}$, so that $|A_p|^2$ and $|A_s|^2$ are 
proportional to the intracavity pump and Stokes energies, whereas $S_{\mathrm{in}}$ has units of 
$\sqrt{\mathrm{power}}$, so that $|S_{\mathrm{in}}|^2$ is proportional to the input pump power. 
Accordingly, the product $\sqrt{\kappa_{ex}}S_{\mathrm{in}}$ has the same units as $\partial_t A_p$ and 
acts as the driving term for the intracavity pump field.

To account for the transverse model structure of the optical and acoustic fields, we 
introduce the overlap factors $\Lambda_F$, $\Lambda_B$, and $\Lambda_\rho$ in the three-wave 
coupling terms. These quantities arise from integrating the full vectorial fields over the 
transverse cross section and effectively reduce the spatially dependent interaction to a 
single-mode description. 
Physically, the overlap factors describe the degree of spatial matching between the pump, 
Stokes, and acoustic modes involved in the Brillouin process. They therefore act as 
multiplicative coefficients that renormalize the coupling strength and define an effective 
nonlinear interaction constant. The explicit expressions of $\Lambda_p$, $\Lambda_s$, and 
$\Lambda_\rho$ are given by Eqs.~(\ref{eq:Lambda_def}), where the transverse field 
profiles are weighted and normalized over the interaction area.
\begin{equation}
\begin{aligned}
\Lambda_p &=
\frac{
\displaystyle \int_A \psi_\rho(x,y)\, \psi_s(x,y)\, \psi_p^*(x,y)\, dA
}{
\displaystyle \int_A \psi_p(x,y)\, \psi_p^*(x,y)\, dA
}
\\[6pt]
\Lambda_s &=
\frac{
\displaystyle \int_A \psi_\rho^*(x,y)\, \psi_p(x,y)\, \psi_s^*(x,y)\, dA
}{
\displaystyle \int_A \psi_s(x,y)\, \psi_s^*(x,y)\, dA
}
\\[6pt]
\Lambda_\rho &=
\frac{
\displaystyle \int_A \psi_p(x,y)\, \psi_s^*(x,y)\, \psi_\rho^*(x,y)\, dA
}{
\displaystyle \int_A \psi_\rho(x,y)\, \psi_\rho^*(x,y)\, dA
}
\end{aligned}
\label{eq:Lambda_def}
\end{equation}
Here, $\psi_p(x,y)$, $\psi_s(x,y)$, and $\psi_\rho(x,y)$
represent the transverse eigenmodes of the pump, Stokes, and acoustic fields.
These mode profiles characterize the spatial distributions in the transverse
plane and are used to evaluate the overlap integrals that determine the
effective three-wave coupling strength.

A representative realization of the continuous-medium picture is the microresonator Brillouin laser 
illustrated in Fig.~\ref{fig:model_compare}(a). In this configuration, the pump field circulates in a 
high-$Q$ microresonator and interacts continuously with the acoustic mode supported by the same structure.
The pump field and the acoustic density wave propagate in the same 
direction to circulate clockwise, while the generated Stokes field propagates in the opposite direction circulating counterclockwise. 
This propagation geometry is required by momentum conservation, \(k_p=k_s+q\), where 
the acoustic wavevector \(q\) compensates the difference between the forward pump momentum and the backward 
Stokes momentum.
Because the pump, Stokes, and acoustic fields
remain co-localized within the same nonlinear medium throughout the cavity evolution, this system is well 
described by the continuous-medium three-wave model.
Within the continuous-medium model, the noise compression of the Stokes field is controlled by the 
competition between the acoustic relaxation rate and the optical decay rate of the Stokes mode. 
Physically, strong acoustic damping acts as a filter for the high-frequency pump phase fluctuations, so that these fluctuations are largely absorbed by the acoustic field and are therefore prevented from being transferred to the Stokes field. As a result, the 
compression factor increases with the ratio of the acoustic damping rate to the Stokes-mode decay 
rate and is given by \cite{Li:12}
\begin{equation}
\label{eq:cm_compression_intro}
\mathcal{C}_{\mathrm{CM}} = 1+\frac{\Gamma_B}{\gamma_s},
\end{equation}
For high-$Q$ microresonators, $\gamma_s \ll \Gamma_B$, so that a large noise-suppression ratio can be achieved.
This mechanism is highly effective in microresonator systems, where the large cavity $Q$ leads to a small 
optical decay rate, although such devices typically operate at relatively low output power. 
In free-space Brillouin lasers, however, power scaling generally requires a lower cavity $Q$ and hence a 
larger optical decay rate than in high-$Q$ microresonator systems \cite{zhang2025frequency}. If the conventional continuous-medium 
model is applied only to the nonlinear gain medium, then the predicted compression factor remains determined 
by the local three-wave interaction inside that medium and does not become exceptionally large under 
realistic free-space operating conditions. In particular, substituting the parameters relevant to the gain 
region alone does not reproduce the strong noise-suppression ratio observed experimentally. This discrepancy 
shows that the free-space propagation outside the gain medium cannot be treated as dynamically irrelevant. 
Instead, it must be incorporated explicitly into the cavity description, since it plays an essential role 
in the noise evolution of the free-space system. This failure motivates a new cavity-level description in 
which the localized Brillouin interaction and the subsequent free-space propagation are treated on an equal 
footing.

\section{Discrete Cavity model}
To capture the essential cavity structure of free-space Brillouin lasers, we introduce a discrete-cavity 
model in which the nonlinear interaction inside the gain medium and the subsequent passive free-space 
propagation are treated as distinct stages of each round trip. Unlike in continuous-medium descriptions, 
the Brillouin interaction is localized within a finite nonlinear medium rather than distributed over 
the entire cavity path. Within this model, the optical field survives over the full cavity round trip,
whereas the acoustic field is generated only inside the gain medium and relaxes during the subsequent 
free-space propagation interval, when no further opto-acoustic interaction occurs. This asymmetry between 
optical storage and acoustic relaxation forms the basis of the discrete-cavity description developed below.
The physical content of the discrete-cavity description can be understood as a temporal separation between 
optical storage and acoustic relaxation introduced by the free-space propagation. As illustrated in 
Fig.~\ref{fig:model_compare}(b), each cavity round trip consists of a short nonlinear interaction inside the gain 
medium followed by a much longer interval of passive free-space propagation. During this passive interval, 
the optical field is largely preserved and continues to circulate in the cavity, whereas the acoustic field 
remains confined to the gain medium and decays in the absence of further driving. Successive Brillouin 
interactions therefore occur in the presence of a partially relaxed acoustic field rather than a continuously 
driven one, as assumed in continuous-medium models. From the viewpoint of the round-trip optical dynamics, 
this interruption of the opto-acoustic coupling appears as an enhanced effective damping rate of the acoustic field. 
As a result, the acoustic field acts as a more effective filter, so that the transfer 
of pump fluctuations to the Stokes field is further suppressed, leading to the large noise-suppression ratio 
in free-space Brillouin lasers.

Inspired by the Ikeda map \cite{ikeda1979multiple}, we formulate the discrete-cavity dynamics, we represent the system by a state vector 
$\mathbf{X}^{(n)}$ that collects the slowly varying amplitudes of the pump field, the 
Stokes field, and the acoustic density wave at the beginning of the $n$-th cavity round trip,
\begin{equation}
\label{state vector}
\mathbf{X}^{(n)}
\equiv
\begin{pmatrix}
A_p^{(n)} \\
A_s^{(n)} \\
\rho^{(n)}
\end{pmatrix}
\end{equation}
The index \(n\) labels successive cavity round trips, while \(A_p^{(n)}\), \(A_s^{(n)}\), 
and \(\rho^{(n)}\) denote the pump field, the Stokes field, and the acoustic density-wave 
amplitude, respectively, immediately before the \(n\)-th interaction event.

Within the discrete-cavity model, the round-trip dynamics naturally decompose into two 
successive stages: the nonlinear Brillouin interaction inside the gain medium and the 
subsequent passive propagation through the free-space cavity. These two processes are 
described by separate evolution operators.
The nonlinear interaction inside the gain medium is represented by an operator  
$\mathcal{U}_{\mathrm{NL}}$, defined as the finite-time evolution generated by the 
three-wave equations (\ref{three_coupled_wave_equations})
over the interaction time in the nonlinear medium. Acting on the state vector, this operator 
produces an intermediate state
\begin{equation}
\mathbf{X}^{(n,+)} = \mathcal{U}_{\mathrm{NL}}\,\mathbf{X}^{(n)} 
\end{equation}
The passive free-space propagation between successive interaction events is described by 
a propagation operator $\mathcal{U}_{\mathrm{free}}$. During this stage no Brillouin 
interaction occurs; the optical fields accumulate phase and experience cavity loss, 
while the acoustic field undergoes free decay governed by its intrinsic damping rate. 
The propagation operator therefore acts diagonally on the state vector,
\begin{equation}
\mathbf{X}^{(n+1)}
=
\mathcal{U}_{\mathrm{free}}\,\mathbf{X}^{(n,+)} 
\end{equation}
with
\begin{equation}
\mathcal{U}_{\mathrm{free}}
=
\begin{pmatrix}
e^{i\phi_{p}} & 0 & 0 \\
0 & e^{i\phi_{s}} & 0 \\
0 & 0 & e^{-\frac{\Gamma_{B}}{2}\tau_{\mathrm{fs}}}
\end{pmatrix}
\end{equation}
Here \(\tau_{\mathrm{fs}}\) denotes the free-space propagation time. The factors
\(e^{i\phi_p}\) and \(e^{i\phi_s}\) represent the phase accumulation of the pump
and Stokes fields during the free-space propagation interval. Since the two
optical fields remain phase locked in the steady operating regime, these
propagation phases do not affect the subsequent coupled dynamics and can be
treated as dynamically irrelevant. By contrast, the acoustic component acquires
the decay factor \(e^{-\frac{\Gamma_B}{2}\tau_{\mathrm{fs}}}\), which describes
the relaxation of the acoustic field during the same interval.
Then the evolution over one cavity round trip is therefore described by the discrete map
\begin{equation}
\label{simulation_eq}
\mathbf{X}^{(n+1)}
=
\left(
\mathcal{U}_{\mathrm{free}}
\circ
\mathcal{U}_{\mathrm{NL}}
\right)
\mathbf{X}^{(n)} 
\end{equation}
which explicitly separates the localized nonlinear interaction from the subsequent 
passive propagation in the cavity.

To obtain an explicit and analytically tractable discrete map for the free-space cavity dynamics, we now 
introduce a delta-kick approximation for the nonlinear evolution inside the gain 
medium. Under this approximation, the Brillouin interaction is treated as a localized event that acts once during 
each cavity round trip, over a finite interaction time $\tau_{\mathrm{int}}$ determined by the transit of 
the optical field through the gain medium. The corresponding fields immediately after the interaction are then obtained 
by integrating the continuous three-wave equations (\ref{delta_kicked_equations}) over a single pass through the medium. To leading order in the interaction time 
$\tau_{\mathrm{int}}$, this procedure yields the following reduced update equations:
\begin{equation}
\begin{aligned}
A_p^{(n+1)}
&=
\left(1-\frac{\gamma}{2}\tau_{\mathrm{int}}\right) A_p^{(n)}
+i g_{p}\,\tau_{\mathrm{int}}\, \rho^{(n)} A_s^{(n)}
+ \sqrt{\kappa_{ex}}\,\tau_{\mathrm{int}}\, S_{\mathrm{in}}
\\
A_s^{(n+1)}
&=
\left(1-\frac{\gamma}{2}\tau_{\mathrm{int}}\right) A_s^{(n)}
+i g_{s}\,\tau_{\mathrm{int}}\, \rho^{*(n)} A_p^{(n)}
\\
\rho^{(n+1)}
&=
\left(1-\frac{\Gamma_B}{2}\tau_{\mathrm{rt}}\right)\rho^{(n)}
+ i g_\rho\,\tau_{\mathrm{int}}\, A_p^{(n)} A_s^{*(n)}
\end{aligned}
\label{delta_kicked_equations}
\end{equation}
The prefactor \(1-\frac{\gamma}{2}\tau_{\mathrm{int}}\) represents the optical survival factor over the short interaction interval \(\tau_{\mathrm{int}}\), obtained to first order from the continuous optical loss term. Here we assume, for simplicity, that the pump and Stokes fields have the same optical decay rate, \(\gamma_p=\gamma_s\equiv\gamma\). By contrast, the factor \(1-\frac{\Gamma_B}{2}\tau_{\mathrm{rt}}\) represents the acoustic survival factor over one cavity round trip, reflecting the fact that the acoustic field continues to decay throughout the full round-trip time \(\tau_{\mathrm{rt}}\), including the passive free-space propagation stage.
Under these assumptions, the effective optical and acoustic coupling strengths are given by
\[
g_{p} \equiv \frac{\omega_{p}\gamma_e}{4n^2\rho_0}\Lambda_{p}
\qquad
g_{s} \equiv \frac{\omega_{s}\gamma_e}{4n^2\rho_0}\Lambda_{s}
\qquad
g_\rho \equiv \frac{\epsilon_0\gamma_e^2 q^2}{4\Omega}\Lambda_{\rho}
\]
The factor \(\tau_{\mathrm{int}}\) appears because these coefficients describe the nonlinear coupling 
accumulated during a single passage through the gain medium, rather than the instantaneous coupling rates 
of the continuous-time equations.
The explicit derivation of these coefficients and of the reduced discrete map is given in 
Appendix~\ref{app:delta_kick_derivation}. Here we only summarize the two approximations underlying 
Eqs.~(\ref{delta_kicked_equations}). First, the continuous three-wave 
interaction is coarse-grained over the finite interaction time $\tau_{\mathrm{int}}$, so that the nonlinear
 coupling is represented as a single kick acting once per cavity round trip. Second, the interaction is 
 assumed to be sufficiently weak that the pump, Stokes, and acoustic amplitudes do not vary appreciably 
 during one pass through the gain medium, which allows the coupling terms to be evaluated using their 
 pre-interaction values.

\section{Steady-state solutions and threshold condition}
\label{Steady-state solutions and threshold condition}
We now analyze the steady-state operation of the discrete-cavity Brillouin laser
described by Eqs.~(\ref{delta_kicked_equations}).
In the discrete-cavity framework, the steady state is defined as a fixed point of
the round-trip map, such that the system state reproduces itself after each cavity
round trip,
\begin{equation}
\begin{aligned}
A_p^{(n+1)} &= A_p^{(n)} \\
A_s^{(n+1)} &= A_s^{(n)} \\
\rho^{(n+1)} &= \rho^{(n)}
\end{aligned}
\end{equation}
Importantly, this condition refers only to the field values evaluated at a specific
reference plane in the cavity. It does not require the fields to remain constant
during the round-trip evolution.
Within each round trip, the optical and acoustic fields may undergo nontrivial
temporal and spatial evolution. The steady state therefore corresponds to a
periodic dynamical state whose values repeat once per cavity cycle.
This notion of steady state naturally arises in the discrete-cavity description,
where the system dynamics are formulated as an iteration of a round-trip map.
The fixed-point condition constrains the fields at the discrete sampling plane
while allowing internal evolution within each cavity cycle.

At steady state, the field amplitudes remain invariant from one
round trip to the next. Substituting the fixed-point condition into
Eqs.~(\ref{delta_kicked_equations}), the lasing threshold is obtained
by considering the onset of a nonzero Stokes solution,
\(A_{s,0}\neq 0\), for which the Brillouin gain exactly compensates
the effective cavity losses. This yields the intracavity pump
threshold intensity
\begin{equation}
\label{ss:pump_threshold}
|A_{p,0}|_{\mathrm{th}}^{2}
=
\frac{\gamma\,\Gamma_B}{4 g_s g_\rho}\,
\frac{\tau_{\mathrm{rt}}}{\tau_{\mathrm{int}}}.
\end{equation}
The corresponding input-pump threshold then follows from the
below-threshold pump relation by setting \(A_{s,0}=0\).
Equation~(\ref{ss:pump_threshold}) shows explicitly that the lasing
threshold scales in proportion to the ratio
\(\tau_{\mathrm{rt}}/\tau_{\mathrm{int}}\). This dependence reflects
the fact that the nonlinear interaction is confined to only a short
segment of each cavity round trip, so that a longer passive
propagation interval requires a correspondingly larger intracavity
pump field to reach threshold.
Above threshold, the intracavity pump field becomes clamped close to
its threshold value. Under this condition, the steady-state Stokes
intensity is given by
\begin{equation}
\label{ss:Stokes_intensity}
|A_{s,0}|^{2}
=
\frac{\Gamma_B}{2 g_s g_\rho}\,
\frac{\tau_{\mathrm{rt}}}{\tau_{\mathrm{int}}}
\left(
\frac{\sqrt{\kappa_{ex}}\,|S_{\mathrm{in}}|}{|A_{p,0}|_{\mathrm{th}}}
-\frac{\gamma}{2}
\right).
\end{equation}
Equation~(\ref{ss:Stokes_intensity}) shows that the Stokes intensity
scales linearly with the effective pump drive above threshold, while
remaining proportional to the ratio \(\tau_{\mathrm{rt}}/\tau_{\mathrm{int}}\).

\begin{figure}[t]
\centering
\includegraphics[width=\columnwidth]{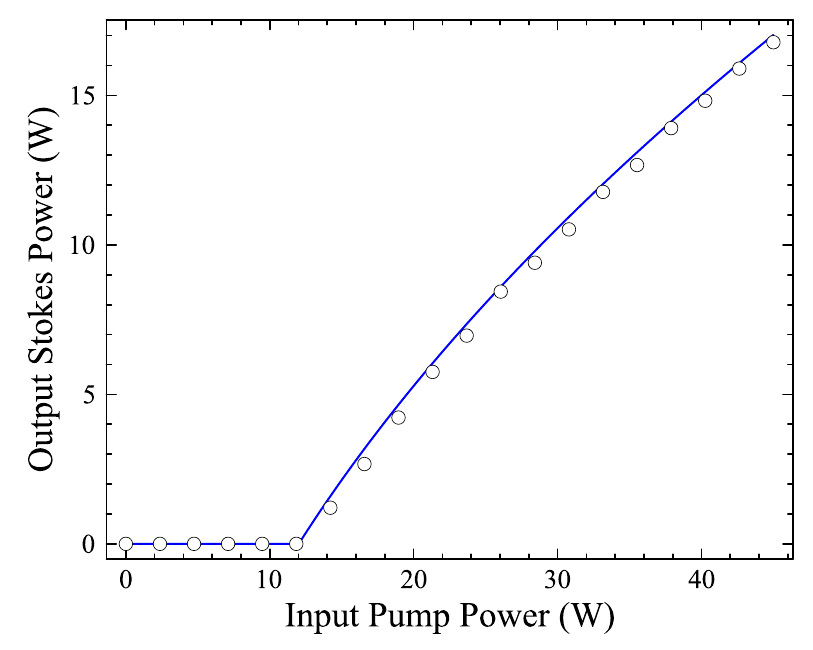}
\caption{
Steady-state output Stokes power as a function of input pump power.
The solid curve represents the analytical steady-state solution of
Eq.~(\ref{ss:Stokes_intensity}), while circles denote numerical
solutions of the discrete cavity map.
The threshold is \(P_{\mathrm{th}}\approx 12~\mathrm{W}\),
above which the Stokes power increases linearly due to pump clamping.
At an input pump power of \(45~\mathrm{W}\), the Stokes output power
reaches approximately \(17~\mathrm{W}\).
}
\label{fig:power_curve}
\end{figure}

To validate the analytical results, we performed time-domain numerical simulations based on the discrete 
round-trip formulation in Eq.s~(\ref{simulation_eq}). The parameters used in the simulations are summarized in 
Table~\ref{tab:parameters}. In each cavity round trip, the nonlinear interaction inside the diamond gain
medium and the subsequent free-space propagation were treated as two separate stages. Within the diamond 
region, the continuous three-wave coupling equations were integrated numerically using a fourth-order 
Runge--Kutta (RK4) scheme, thereby resolving the nonlinear optical--acoustic interaction over the finite 
interaction time $\tau_{\mathrm{int}}$. The remaining free-space segment was then implemented as a linear 
propagation step, during which the optical fields evolved passively and the acoustic field continued to 
decay over the rest of the round-trip time. At each round trip, the complex pump, Stokes, and acoustic 
amplitudes were recorded to extract the intracavity intensities and phase trajectories. The steady-state 
output power and the corresponding noise spectra were then evaluated from the simulated time-domain signals.
Figure~\ref{fig:power_curve} shows the steady-state output Stokes power as a function of the input pump power.
The solid blue curve represents the analytical prediction obtained from
Eq.~(\ref{ss:Stokes_intensity}), while the open circles denote the numerical
solutions of the discrete round-trip map described by Eq.~(\ref{simulation_eq}).
A clear lasing threshold is observed at $P_{\mathrm{th}}\approx 12~\mathrm{W}$,
above which the Stokes power increases monotonically with the excess pump.
The excellent agreement between the analytical curve and the numerical data
confirms the validity of the steady-state approximation and the pump-clamping
assumption employed in the theoretical derivation.
For the parameter set used here, an input pump power of $45~\mathrm{W}$
yields an output Stokes power of approximately $17~\mathrm{W}$,
in agreement with the experimentally reported power level\cite{zhang2025frequency}, whose 
Stokes output power is $14.6~\mathrm{W}$ as a pump power of $45~\mathrm{W}$.
\begin{table}[t]
\caption{Characteristic parameters of the free-space diamond Brillouin laser used in the simulations.}
\label{tab:parameters}
\begin{ruledtabular}
\begin{tabular}{lc}
Parameter & Value \\
\hline
Crystal length $L_c$ (m) & $5.0\times10^{-3}$ \\
Cavity length $L$ (m) & $5.08\times10^{-1}$ \\
Refractive index $n$ & 2.39 \\
Acoustic damping rate $\Gamma_B$ (MHz) & $2\pi\times 5.3$ \\
Electrostrictive constant $\gamma_e$ & 8.16 \\
Pump wavelength $\lambda_p$ (nm) & 1064 \\
Mass density $\rho_0$ (g/cm$^{3}$) & 3.515 \\
Acoustic velocity $v$ (km/s) & 18 \\
Beam waist $w_0$ ($\mu$m) & 74 \\
External coupling rate $\kappa_{\mathrm{ex}}$ (s$^{-1}$) & $2.3\times10^{7}$ \\
Optical cavity decay rate $\gamma$ (s$^{-1}$) & $3.6\times10^{7}$ \\
Interaction time $\tau_{\mathrm{int}}$ (s) & $4.0\times10^{-11}$ \\
Round-trip time $\tau_{\mathrm{rt}}$ (s) & $1.7\times10^{-9}$ \\
Acoustic wavevector $q$ (m$^{-1}$) & $2.8\times10^{7}$ \\
Acoustic angular frequency $\Omega$ (s$^{-1}$) & $5.1\times10^{11}$ \\
Optical mode volume $V_{\mathrm{opt}}$ (m$^{3}$) & $8.6\times10^{-11}$ \\
\end{tabular}
\end{ruledtabular}
\end{table}
\section{Noise Dynamics}
\subsection{Noise compression and scaling behavior}
To analyze the frequency-domain response of the discrete-cavity
Brillouin laser, we now perform a small-signal analysis around the
steady-state operating point obtained in Sec.~IV. This approach allows
us to quantify how weak fluctuations evolve from round trip to round
trip and how pump noise is transferred, filtered, or suppressed by the
coupled optical--acoustic dynamics. We therefore linearize the
round-trip evolution equations, Eq.~(\ref{delta_kicked_equations}), about the steady
state and introduce small perturbations of the pump, Stokes, and
acoustic fields as
\begin{equation}
\begin{aligned}
A_p^{(n)} &= A_{p,0} + \delta A_p^{(n)} \\
A_s^{(n)} &= A_{s,0} + \delta A_s^{(n)} \\
\rho^{(n)} &= \rho'_0 + \delta \rho^{(n)} 
\end{aligned}
\label{lin:expansion}
\end{equation}
where \(\delta A_p^{(n)}\), \(\delta A_s^{(n)}\), and \(\delta \rho^{(n)}\)
represent small perturbations around the steady state. These quantities
are generally complex and describe both amplitude and phase fluctuations
of the fields.
Substituting these expressions into
Eqs.~(\ref{delta_kicked_equations})
and retaining only first-order terms yields the linearized evolution equations.

\begin{equation}
\begin{aligned}
\delta A_{p}^{(n+1)}
&=
\left(1-\frac{\gamma}{2}\tau_{\mathrm{int}}\right)\delta A_{p}^{(n)}
\\
&\quad
+i g_p\,\tau_{\mathrm{int}}
\left(
\rho_{0}\,\delta A_{s}^{(n)}
+
A_{s,0}\,\delta \rho^{(n)}
\right)
\\
&\quad
+\sqrt{\kappa_{\mathrm{ex}}}\,\tau_{\mathrm{int}}\,\delta S_{\mathrm{in}}^{(n)}
\\[4pt]
\delta A_{s}^{(n+1)}
&=
\left(1-\frac{\gamma}{2}\tau_{\mathrm{int}}\right)\delta A_{s}^{(n)}
\\
&\quad
+i g_s\,\tau_{\mathrm{int}}
\left(
\rho_{0}^{*}\,\delta A_{p}^{(n)}
+
A_{p,0}\,\delta \rho^{*(n)}
\right)
\\[4pt]
\delta \rho^{(n+1)}
&=
\left(1-\frac{\Gamma_{B}}{2}\tau_{\mathrm{rt}}\right)\delta \rho^{(n)}
\\
&\quad
+i g_{\rho}\,\tau_{\mathrm{int}}
\left(
A_{s,0}^{*}\,\delta A_{p}^{(n)}
+
A_{p,0}\,\delta A_{s}^{*(n)}
\right)
\end{aligned}
\label{lin:fluct}
\end{equation}
At this stage, we consider the cavity response to fluctuations of the input pump field, represented 
by \(\delta S_{\mathrm{in}}^{(n)}\). To isolate this externally driven noise transfer, we temporarily 
neglect the thermal Langevin force acting on the acoustic field. We first focus on the phase-noise 
response. In the above-threshold steady state, the pump field is clamped near its threshold value and 
the stationary solution is phase locked, so that amplitude fluctuations contribute only as higher-order 
corrections to the phase dynamics.
We therefore neglect amplitude fluctuations at this stage and express the optical, acoustic, and input-field fluctuations in terms of phase perturbations as
\begin{equation}
\begin{aligned}
\delta A_p^{(n)} &\simeq i A_{p,0}\,\delta\varphi_p^{(n)}, \\
\delta A_s^{(n)} &\simeq i A_{s,0}\,\delta\varphi_s^{(n)}, \\
\delta \rho^{(n)} &\simeq i \rho'_{0}\,\delta\varphi_\rho^{(n)}, \\
\delta S_{\mathrm{in}}^{(n)} &\simeq i S_{\mathrm{in},0}\,\delta\varphi_{\mathrm{in}}^{(n)}.
\end{aligned}
\end{equation}
Here \(\delta\varphi_{p,n}\), \(\delta\varphi_{s,n}\), \(\delta\varphi_{\rho,n}\), and 
\(\delta\varphi_{\mathrm{in},n}\) denote the phase perturbations of the intracavity pump, Stokes, acoustic, 
and input pump fields, respectively. Under the phase-only approximation, the fluctuation dynamics are reduced to the coupled evolution of the pump, Stokes, and acoustic phases. Physically, the pump phase acts as the external driving variable, while the Stokes and acoustic phases are coupled through the discrete Brillouin interaction and the round-trip acoustic relaxation. Using the steady-state relations together with the threshold condition derived in Sec.~\ref{Steady-state solutions and threshold condition}, the nonlinear coupling terms can be rewritten in terms of steady-state cavity parameters. The resulting phase recursions are
\begin{equation}
\begin{aligned}
\delta\varphi_{p}^{(n+1)}
&=
\left(1-\frac{\gamma}{2}\tau_{\mathrm{int}}\right)\delta\varphi_{p}^{(n)}
-\frac{\gamma}{2}\tau_{\mathrm{int}}
\frac{|A_{s,0}|^2}{|A_{p,0}|_{\mathrm{th}}^2}
\left(
\delta\varphi_{s}^{(n)}+\delta\varphi_{\rho}^{(n)}
\right)
\\
&\quad+
\frac{\gamma}{2}\tau_{\mathrm{int}}
\left(
1+\frac{|A_{s,0}|^2}{|A_{p,0}|_{\mathrm{th}}^2}
\right)
\delta\varphi_{\mathrm{in}}^{(n)},
\\[4pt]
\delta\varphi_{s}^{(n+1)}
&=
\left(1-\frac{\gamma}{2}\tau_{\mathrm{int}}\right)\delta\varphi_{s}^{(n)}
+
\frac{\gamma}{2}\tau_{\mathrm{int}}
\left(
\delta\varphi_{p}^{(n)}-\delta\varphi_{\rho}^{(n)}
\right),
\\[4pt]
\delta\varphi_{\rho}^{(n+1)}
&=
\left(1-\frac{\Gamma_{B}}{2}\tau_{\mathrm{rt}}\right)\delta\varphi_{\rho}^{(n)}
+
\frac{\Gamma_{B}}{2}\tau_{\mathrm{rt}}
\left(
\delta\varphi_{p}^{(n)}-\delta\varphi_{s}^{(n)}
\right).
\end{aligned}
\label{eq:phase_recursions_full}
\end{equation}
Equation~(\ref{eq:phase_recursions_full}) provides the discrete-time phase dynamics of the pump, Stokes, and 
acoustic fields under external pump-phase driving.
\begin{figure*}[htbp]
\centering
\begin{subfigure}{0.49\textwidth}
\centering
\includegraphics[width=\linewidth]{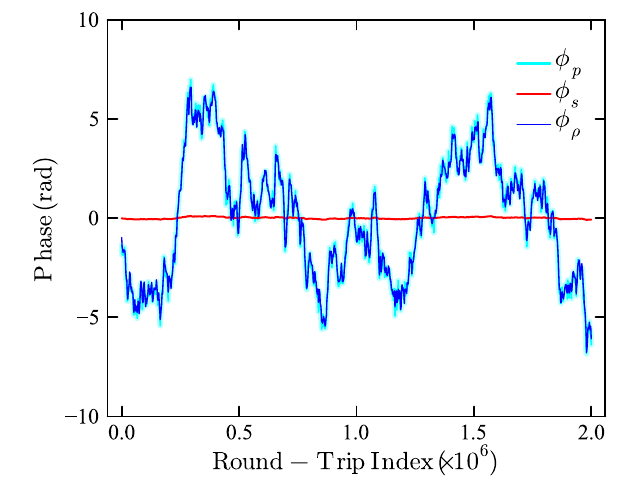}
\end{subfigure}
\hfill
\begin{subfigure}{0.49\textwidth}
\centering
\includegraphics[width=\linewidth]{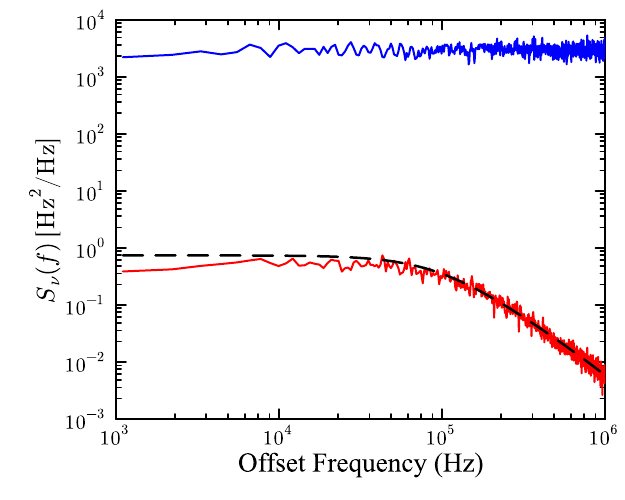}
\end{subfigure}
\caption{
(a) Simulated phase evolution versus round-trip index.The blue curve represents the pump phase, while the red curve denotes the Stokes phase.
(b) Frequency-noise spectra from numerical simulation.
Pump noise (blue), Stokes noise (red), and analytical prediction
(black dashed).}
\label{phase}
\end{figure*}

To analyze the phase-noise transfer in the frequency domain, we apply the discrete-time Fourier transform \cite{5311871}
to the round-trip-sampled phase fluctuations. For a generic sequence \(\varphi^{(n)}\), we define
\begin{equation}
\label{z}
\Phi(\omega)=\sum_{n=-\infty}^{\infty}\varphi^{(n)}e^{-i\omega n\tau_{\mathrm{rt}}}
\end{equation}
where \(\tau_{\mathrm{rt}}\) is the cavity round-trip time. Under this transform, a one-step shift in the 
round-trip index corresponds to multiplication by \(e^{i\omega\tau_{\mathrm{rt}}}\). Neglecting transient 
contributions associated with the initial conditions, the phase recursion relations in 
Eq.~(\ref{eq:phase_recursions_full}) are converted into algebraic equations for the Fourier-domain 
phase fluctuations \(\Phi_p(z)\), \(\Phi_s(z)\), \(\Phi_{\rho}(z)\), and \(\Phi_{\mathrm{in}}
(z)\).
Using the discrete-time phase recursion relations, the acoustic phase can be eliminated to obtain the transfer function from the input pump phase to the Stokes phase,
\begin{widetext}
\begin{equation}
H_{\mathrm{in}\to s}(z)
=
\frac{\Phi_s(z)}{\Phi_{\mathrm{in}}(z)}
=
\frac{
\left(\dfrac{\gamma}{2}\tau_{\mathrm{int}}\right)^2
\left(1+\dfrac{|A_{s,0}|^2}{|A_{p,0}|_{\mathrm{th}}^2}\right)
}{
(z-1)^2
+
\left(\gamma\tau_{\mathrm{int}}+\dfrac{\Gamma_B}{2}\tau_{\mathrm{rt}}\right)(z-1)
+
\dfrac{\gamma}{2}\tau_{\mathrm{int}}
\left(\dfrac{\gamma}{2}\tau_{\mathrm{int}}+\dfrac{\Gamma_B}{2}\tau_{\mathrm{rt}}\right)
\left(1+\dfrac{|A_{s,0}|^2}{|A_{p,0}|_{\mathrm{th}}^2}\right)
}
\label{Hins}
\end{equation}
\end{widetext}
The transfer function \(H_{\mathrm{in}\to s}(z)\) characterizes the
frequency-dependent transfer of pump phase fluctuations to the
Stokes phase. Here the complex variable \(z\) is associated with the
discrete-time Fourier transform defined in Eq.~(\ref{z}). Since the index
\(n\) labels successive cavity round trips separated by the round-trip
time \(\tau_{\mathrm{rt}}\), a harmonic phase fluctuation at Fourier
frequency \(\omega\) corresponds to the identification
\begin{equation}
z = e^{i\omega\tau_{\mathrm{rt}}}
\end{equation}
The phase transfer function of Eq.s~(\ref{Hins}) derived above determines how pump phase
fluctuations are filtered by the discrete cavity dynamics and
transferred to the Stokes field. However, in laser noise analysis the
quantity of direct physical relevance is the frequency fluctuation,
since the instantaneous optical frequency is defined as the time
derivative of the optical phase. Accordingly, a phase fluctuation
\(\varphi(t)\) corresponds to a frequency fluctuation \cite{DiDomenico:10}
\begin{equation}
\delta\nu(t)=\frac{1}{2\pi}\frac{d\varphi(t)}{dt}.
\end{equation}
In the frequency domain, this relation implies that phase and frequency
noise spectra are directly related through a multiplicative factor of
\(\omega^2\). Denoting by \(S_{\varphi}(\omega)\) the phase-noise
power spectral density, the corresponding frequency-noise spectrum is
given by \cite{DiDomenico:10}
\begin{equation}
S_{\nu}(\omega)
=
\left(\frac{\omega}{2\pi}\right)^2
S_{\varphi}(\omega).
\end{equation}
Therefore, once the phase-noise transfer function is known, the
frequency-noise spectrum of the Stokes field can be obtained by
multiplying the input pump phase-noise spectrum by the squared modulus
of the transfer function and then applying the phase-to-frequency
conversion described above.

In experiments, the noise properties of the pump laser are typically
characterized by its measured frequency-noise power spectral density.
Over the frequency range of interest, the pump frequency noise can often
be approximated as white noise, corresponding to a flat spectrum in the
frequency domain.
Using the transfer function derived above, the frequency-noise spectrum
of the Stokes field can be obtained directly from the pump noise
spectrum. In the frequency domain, the Stokes noise spectrum is related
to the pump noise spectrum through
\begin{equation}
S_{\nu_s}(\omega)
=
\left|
H_{\mathrm{in}\rightarrow s}(e^{i\omega\tau_{\mathrm{rt}}})
\right|^2
\,S_{\nu_{\mathrm{in}}}(\omega),
\label{eq:Stokes_noise_transfer}
\end{equation}
where \(S_{\nu_{\mathrm{in}}}(\omega)\) denotes the measured
frequency-noise spectrum of the input pump field. Equation
(\ref{eq:Stokes_noise_transfer}) therefore predicts the Stokes noise
spectrum directly from the experimentally known pump noise spectrum and
the cavity transfer function.

To quantify the noise suppression provided by the Brillouin laser,
it is convenient to introduce the ratio between the input pump noise
spectrum and the resulting Stokes noise spectrum. Physically, this
ratio measures how much of the original pump noise is suppressed
during the nonlinear conversion process inside the cavity.
Accordingly, we define the noise-suppression ratio as
\begin{equation}
\mathcal{C}(\omega)
\equiv
\frac{S_{\nu_{\mathrm{in}}}(\omega)}{S_{\nu_s}(\omega)}=\frac{1}{
\left|H_{\mathrm{in}\rightarrow s}
\!\left(e^{i\omega\tau_{\mathrm{rt}}}\right)\right|^2
}.
\label{eq:suppression_ratio}
\end{equation}
where \(S_{\nu_{\mathrm{in}}}(\omega)\) and \(S_{\nu_s}(\omega)\)
denote the frequency-noise power spectral densities of the input pump
and the generated Stokes field, respectively. 
In the low-frequency limit \(\omega\to 0\), the noise-suppression ratio reduces to
\begin{equation}
\mathcal{C}(0)
=
\left(
1+\frac{\Gamma_B\,\tau_{\mathrm{rt}}}{\gamma\,\tau_{\mathrm{int}}}
\right)^2 .
\label{eq:R0}
\end{equation}
Equation~(\ref{eq:R0}) shows explicitly that the low-frequency noise
suppression is not determined by the material parameters alone, but by
the ratio between the full cavity round-trip time and the short
interaction time inside the gain medium. This result highlights the
essential role of the free-space propagation segment differnet from the Eq.s~(\ref{eq:cm_compression_intro})
of the continnum model.
This is due to the acoustic field
continues to decay over the entire round-trip time \(\tau_{\mathrm{rt}}\),
whereas the nonlinear optical interaction is accumulated only over the
much shorter interval \(\tau_{\mathrm{int}}\). Therefore, the unusually
large noise-suppression ratios observed in free-space Brillouin-laser
experiments can only be explained when the passive free-space
propagation is incorporated explicitly into the cavity dynamics.

To further validate the analytical predictions, we performed time-domain simulations of the discrete-cavity 
dynamics using the parameters listed in Table~\ref{tab:parameters}. In the simulation, the pump laser was 
modeled with ideal white frequency noise corresponding to a Lorentzian linewidth of \(3~\mathrm{kHz}\), 
and the resulting stochastic phase fluctuations were injected into the input pump field. The coupled optical 
and acoustic evolution was then computed over successive cavity round trips using the discrete-cavity map.
The left panel of Fig.~\ref{phase} shows the simulated phase trajectories. The pump phase exhibits large 
random excursions characteristic of a diffusive phase process, whereas the Stokes phase remains strongly 
confined around zero with only small residual fluctuations. This strong contrast directly reflects the large 
phase-noise suppression achieved in the discrete cavity. At the same time, the acoustic phase closely follows 
the pump phase fluctuations, indicating that a substantial portion of the pump phase noise is transferred 
into the acoustic degree of freedom rather than into the Stokes field.
A more complete validation is provided by the frequency-noise spectra shown in the right panel. The blue 
curve denotes the simulated pump frequency-noise spectrum, which remains nearly flat over the analyzed 
offset-frequency range, consistent with the imposed white frequency-noise model. The red curve shows 
the corresponding Stokes frequency-noise spectrum, exhibiting a pronounced low-frequency suppression.
The black dashed line represents the analytical prediction obtained from the compression factor derived 
above. The good agreement between the simulated and analytical spectra confirms that the discrete-cavity 
theory correctly captures the phase-noise transfer dynamics and explains the large suppression ratio 
observed in free-space Brillouin lasers.
This prediction is also consistent with the
experimental observation that the Stokes frequency-noise spectrum remains
at a level of approximately \(3~\mathrm{Hz}^2/\mathrm{Hz}\) over the
offset-frequency range from \(10^3\) to \(10^6~\mathrm{Hz}\) \cite{zhang2025frequency}.

\subsection{Fundmental linewidth from thermal noise}
So far, we have considered the deterministic cavity response and the transfer of external pump noise in the 
absence of internal stochastic driving. To determine the fundamental noise limit of the Brillouin laser, it 
is necessary to include the thermal fluctuations of the acoustic density wave explicitly. Because the 
acoustic mode is coupled to a finite-temperature reservoir, it is subject not only to deterministic damping 
but also to random Langevin forcing. This thermally driven acoustic fluctuation persists even in the absence 
of technical pump noise and therefore sets the fundamental floor for the Stokes linewidth and relative 
intensity noise.

To evaluate the intrinsic linewidth, we now consider the thermally driven phase dynamics of the discrete 
cavity. Compared with the phase-recursion equations (\ref{eq:phase_recursions_full}) used above for pump-noise transfer, the present 
formulation removes the external input pump-phase fluctuation as the driving term and instead introduces a stochastic Langevin contribution in the acoustic equation. In this way, the deterministic coupling between the pump, Stokes, and acoustic phases remains unchanged, while the only stochastic drive is provided by the thermally excited acoustic field. Linearizing the stochastic discrete-cavity equations around the steady-state operating point and retaining only phase fluctuations to leading order, we obtain the following recursion relations.
\begin{equation}
\begin{aligned}
\delta\varphi_{p}^{(n+1)}
&=
\left(1-\frac{\gamma}{2}\tau_{\mathrm{int}}\right)\delta\varphi_{p}^{(n)}
\\
&\quad
-\frac{\gamma}{2}\tau_{\mathrm{int}}
\frac{|A_{s,0}|^2}{|A_{p,0}|_{\mathrm{th}}^2}
\left(
\delta\varphi_{s}^{(n)}+\delta\varphi_{\rho}^{(n)}
\right)
\\[6pt]
\delta\varphi_{s}^{(n+1)}
&=
\left(1-\frac{\gamma}{2}\tau_{\mathrm{int}}\right)\delta\varphi_{s}^{(n)}
+\frac{\gamma}{2}\tau_{\mathrm{int}}
\left(
\delta\varphi_{p}^{(n)}-\delta\varphi_{\rho}^{(n)}
\right)
\\[6pt]
\delta\varphi_{\rho}^{(n+1)}
&=
\left(1-\frac{\Gamma_{B}}{2}\tau_{\mathrm{rt}}\right)\delta\varphi_{\rho}^{(n)}
\\
&\quad
+\frac{\Gamma_{B}}{2}\tau_{\mathrm{rt}}
\left(
\delta\varphi_{p}^{(n)}-\delta\varphi_{s}^{(n)}
\right)
+\eta^{(n)}
\end{aligned}
\label{eq:intrinsic_phase_recursions}
\end{equation}
Here \(\eta^{(n)}\) denotes the stochastic phase increment induced by the thermal Langevin force acting on the acoustic mode.
The corresponding discrete acoustic fluctuation over one round trip
is obtained by integrating the continuous noise force,
\begin{equation}
\label{Langevin}
F_\rho^{(n)}
=
\int_{n\tau_{\mathrm{rt}}}^{(n+1)\tau_{\mathrm{rt}}}
f_B(t)\, dt
\end{equation}
where $f_B(t)$ is the complex white-noise force introduced in
Appendix~\ref{app:Langevin_noise}. The phase noise entering the
linearized phase dynamics is obtained by projecting this complex
increment onto the acoustic phase quadrature,
\begin{equation}
\eta^{(n)}
=
\frac{\mathrm{Im}\!\left[F_\rho^{(n)}\,\rho_0^{'*}\right]}
{|\rho'_0|^2}
\label{eq:eta_def_clean}
\end{equation}
so that only the component of the fluctuation orthogonal to the
steady-state acoustic amplitude contributes to phase diffusion.

Following the same procedure as in the pump-noise transfer analysis, we apply the discrete-time Fourier 
transform defined in Eq.~(\ref{z}) to the stochastic phase recursion relations in 
Eq.~(\ref{eq:intrinsic_phase_recursions}) and neglect the transient contributions associated with the 
initial conditions. This converts the discrete recursion relations into algebraic equations in the 
frequency domain. Eliminating the pump and acoustic phase variables then yields the transfer relation 
from the thermal Langevin increment \(\eta\) to the Stokes phase fluctuation, which can be written as
\begin{widetext}
\begin{equation}
H_{\eta\to s}(z)
\equiv
\frac{\Phi_s(z)}{\eta(z)}
=
-\frac{
\dfrac{\gamma}{2}\tau_{\mathrm{int}}
\left[
(z-1)
+
\dfrac{\gamma}{2}\tau_{\mathrm{int}}
\left(
1+\dfrac{|A_{s,0}|^2}{|A_{p,0}|_{\mathrm{th}}^2}
\right)
\right]
}{
(z-1)
\left[
(z-1)^2
+
\left(
\gamma\tau_{\mathrm{int}}
+\dfrac{\Gamma_B}{2}\tau_{\mathrm{rt}}
\right)(z-1)
+
\dfrac{\gamma}{2}\tau_{\mathrm{int}}
\left(
\dfrac{\gamma}{2}\tau_{\mathrm{int}}
+\dfrac{\Gamma_B}{2}\tau_{\mathrm{rt}}
\right)
\left(
1+\dfrac{|A_{s,0}|^2}{|A_{p,0}|_{\mathrm{th}}^2}
\right)
\right]
}
\label{eq:Heta_s}
\end{equation}
\end{widetext}

The Stokes frequency-noise spectrum generated by the thermal Langevin force is therefore given by
\begin{equation}
S_{\nu_s}(\omega)
=
\left(\frac{\omega}{2\pi}\right)^2
\left|
H_{\eta\to s}\!\left(e^{i\omega\tau_{\mathrm{rt}}}\right)
\right|^2
S_{\eta}(\omega),
\label{eq:Snu_eta}
\end{equation}
where \(S_{\eta}(\omega)\) denotes the power spectral density of the acoustic Langevin phase increment.
The factor \((z-1)^{-1}\) in \(H_{\eta\to s}(z)\) reflects the diffusive accumulation of phase fluctuations 
and therefore corresponds to the usual phase-diffusion pole at low frequency. This apparent singular 
behavior is removed when passing from phase noise to frequency noise, because the time derivative 
relating frequency to phase contributes a compensating factor proportional to \((z-1)\) in the 
low-frequency limit. As a result, the Stokes frequency-noise spectrum remains finite as \(\omega\to 0\), and 
its zero-frequency value determines the fundamental white frequency-noise floor.
And the discrete acoustic phase noise corresponds to a
white-noise process with spectral density
\begin{equation}
S_{\eta}(\omega)
=\left\langle \eta^{(n)}\eta^{(n)}\right\rangle
=
\frac{\tau_{rt} C_B}{2|\rho'_0|^2},
\label{eq:Seta_def_clean}
\end{equation}
where $C_B$ denotes the discrete acoustic Langevin noise strength,
which in the weak-damping limit can be written as
\begin{equation}
C_B \simeq
\frac{k_B T \rho_0 \Gamma_B}{v^2 V_{\mathrm{opt}}}\,
\tau_{\mathrm{int}} .
\end{equation}
Here $k_B$ is the Boltzmann constant, $T$ is the temperature,
$V_{\mathrm{opt}}$ denotes the effective optical mode volume
participating in the Brillouin interaction. 
This expression follows from the fluctuation--dissipation \cite{PhysRevA.42.5514}
relation for thermally driven acoustic fluctuations (see
Appendix~\ref{app:Langevin_noise}).

Because the intrinsic linewidth is set by the white low-frequency floor of the frequency-noise spectrum, we evaluate the zero-frequency limit of Eq.~(\ref{eq:Snu_eta}). In this limit, the phase-diffusion pole contained in the Langevin transfer function is canceled by the phase-to-frequency conversion, leaving a finite white frequency-noise plateau. The zero-frequency Stokes frequency-noise floor is then obtained as
\begin{equation}
S_{\nu,s}(0)
=
\frac{1}{(2\pi)^2\tau_{\mathrm{rt}}^2}
\,
\frac{(\gamma\tau_{\mathrm{int}})^2}
{\left(\gamma\tau_{\mathrm{int}}+\Gamma_B\tau_{\mathrm{rt}}\right)^2}
\,
S_{\eta}(0).
\label{eq:Sv_zero}
\end{equation}
For white frequency noise, the intrinsic linewidth is directly determined by this low-frequency plateau \cite{DiDomenico:10}. 
Using the acoustic Langevin noise spectrum defined above Eq(\ref{eq:Seta_def_clean}), we obtain
\begin{equation}
\Delta\nu_{\mathrm{int}}
=
\frac{C_B}{8\pi^{2}\,|\rho'_0|^2\,\tau_{\mathrm{rt}}}
\,
\frac{1}
{\left(1+\dfrac{\Gamma_B\tau_{\mathrm{rt}}}{\gamma\tau_{\mathrm{int}}}\right)^2}
\label{eq:linewidth_explicit_corrected}
\end{equation}

\begin{figure}[t]
\centering
\includegraphics[width=\columnwidth]{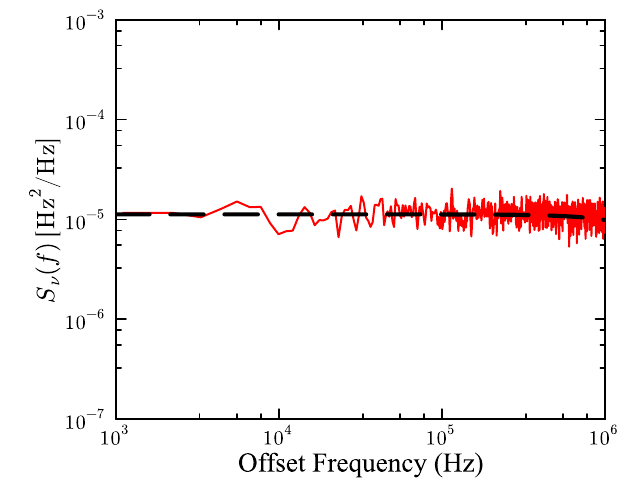}
\caption{Simulated fundamental frequency-noise spectrum of the Stokes field (red). The black dashed curve shows the theoretical prediction.
}
\label{fundamental}
\end{figure}

To verify the analytical prediction for the intrinsic noise floor,
we performed numerical simulations including the thermal Langevin
force acting on the acoustic mode. The stochastic increment
$\eta^{(n)}$ was activated in the acoustic phase equation,
representing thermally drive acoustic fluctuations in the absence
of external pump noise.
From the simulated phase trajectories, the corresponding
frequency-noise spectrum of the Stokes field was extracted.
As shown in Fig.~\ref{fundamental}, the red curve represents the
simulated spectrum, while the black dashed line denotes the
analytical prediction of Eq.~(\ref{eq:Sv_zero}). The spectrum
exhibits a flat low-frequency plateau characteristic of white
frequency noise, which directly determines the intrinsic
linewidth. The excellent agreement between the simulated noise
floor and the theoretical prediction confirms the validity of
the analytical model for the fundamental noise limit.

\subsection{Relative intensity noise}
We now analyze the relative intensity noise (RIN) of the Stokes field
within the discrete-cavity model. In contrast to the previous
phase-noise analysis, we here consider the deterministic amplitude
response of the cavity to pump intensity modulation. To isolate this
mechanism, the thermal Langevin force acting on the acoustic mode is
set to zero and pump phase fluctuations are neglected, so that the
pump intensity modulation remains the only external driving source.
We linearize the dynamical equations around the steady-state solution
and retain only amplitude perturbations. The optical and acoustic
fields, together with the input pump field, are therefore written as
\begin{equation}
\begin{aligned}
A_p^{(n)} &= A_{p,0}\left(1+\delta u_p^{(n)}\right), \\
A_s^{(n)} &= A_{s,0}\left(1+\delta u_s^{(n)}\right), \\
\rho^{(n)} &= \rho_0\left(1+\delta u_\rho^{(n)}\right), \\
S_{\mathrm{in}}^{(n)} &= S_{\mathrm{in},0}\left(1+\delta u_{S_{\mathrm{in}}}^{(n)}\right).
\end{aligned}
\end{equation}
where \(\delta u_p^{(n)}\), \(\delta u_s^{(n)}\), \(\delta u_\rho^{(n)}\), 
and \(\delta u_{S_{\mathrm{in}}}^{(n)}\) denote small real amplitude perturbations 
about the steady-state values.

Since the relative intensity noise is defined in terms of normalized
intensity fluctuations, the RIN transfer function is naturally defined
as the ratio between the normalized Stokes intensity perturbation and
the normalized input pump intensity perturbation. For pure amplitude
perturbations, one has \(\delta I_s/I_{s,0}\simeq 2\delta u_s^{(n)}\) and
\(\delta I_{\mathrm{in}}/I_{\mathrm{in},0}\simeq 2\delta u_{S_{\mathrm{in}}}^{(n)}\),
so that the corresponding RIN transfer function reduces to
\begin{equation}
H_{\mathrm{RIN}}(z)
\equiv
\frac{\delta u_s(z)}{\delta u_{S_{\mathrm{in}}}^{(n)}(z)}
\end{equation}
Substituting the perturbative expansions into the linearized
round-trip equations and transforming the system into the
$z$ domain yields a set of algebraic relations describing
the discrete-cavity response to pump intensity modulation.
Solving these equations for $\delta u_s(z)$ in terms of the
input fluctuation $\delta u_{S_{\mathrm{in}}}^{(n)}(z)$ yields the RIN
transfer function
\begin{widetext}
\begin{equation}
H_{\mathrm{RIN}}(z)
=
\frac{
\left(\dfrac{\gamma}{2}\tau_{\mathrm{int}}\right)^2
\left(1+\dfrac{|A_{s,0}|^2}{|A_{p,0}|_{\mathrm{th}}^2}\right)
\left(
\Delta_b+\dfrac{\Gamma_B}{2}\tau_{\mathrm{rt}}
\right)
}{
\Delta_b\Delta_p^2
+
\left(\dfrac{\gamma}{2}\tau_{\mathrm{int}}\right)
\left(\dfrac{\Gamma_B}{2}\tau_{\mathrm{rt}}\right)
\dfrac{|A_{s,0}|^2}{|A_{p,0}|_{\mathrm{th}}^2}\,
\Delta_p
+
\left(\dfrac{\gamma}{2}\tau_{\mathrm{int}}\right)^2
\dfrac{|A_{s,0}|^2}{|A_{p,0}|_{\mathrm{th}}^2}\,
\Delta_b
+
2\left(\dfrac{\gamma}{2}\tau_{\mathrm{int}}\right)^2
\left(\dfrac{\Gamma_B}{2}\tau_{\mathrm{rt}}\right)
\dfrac{|A_{s,0}|^2}{|A_{p,0}|_{\mathrm{th}}^2}
}.
\label{rin}
\end{equation}
\end{widetext}
where the shorthand quantities are defined as
\begin{equation}
\begin{aligned}
\Delta_p &= z - \left(1-\frac{\gamma}{2}\tau_{\mathrm{int}}\right) \\
\Delta_b &= z - \left(1-\frac{\Gamma_B}{2}\tau_{\mathrm{rt}}\right)
\end{aligned}
\end{equation}

Then RIN spectrum of the Stokes field can then be obtained
directly from the transfer function. For a linear system,
the output noise spectrum is given by the input spectrum
multiplied by the squared magnitude of the transfer function.
Accordingly, the Stokes RIN spectrum reads
\begin{equation}
\mathrm{RIN}_s(\omega)
=
\left|H_{\mathrm{RIN}}(\omega)\right|^2
\,\mathrm{RIN}_{\mathrm{in}}(\omega).
\end{equation}
To characterize the low-frequency behavior of the intensity noise,
we consider the zero-frequency limit of the transfer function.
Substituting \( z \to 1 \) into Eq.~(\ref{rin}) yields the
low-frequency RIN transfer factor
\begin{equation}
|H_{\mathrm{RIN}}(1)|^2
=
\left|
\frac{
2\left(1+\dfrac{|A_{s,0}|^2}{|A_{p,0}|_{\mathrm{th}}^2}\right)
}{
1+4\dfrac{|A_{s,0}|^2}{|A_{p,0}|_{\mathrm{th}}^2}
}
\right|^2.
\end{equation}
This expression shows that the low-frequency RIN transfer remains of order unity, 
varying between 2 at threshold and 1/2 far above threshold.
\begin{figure}[t]
\centering
\includegraphics[width=\columnwidth]{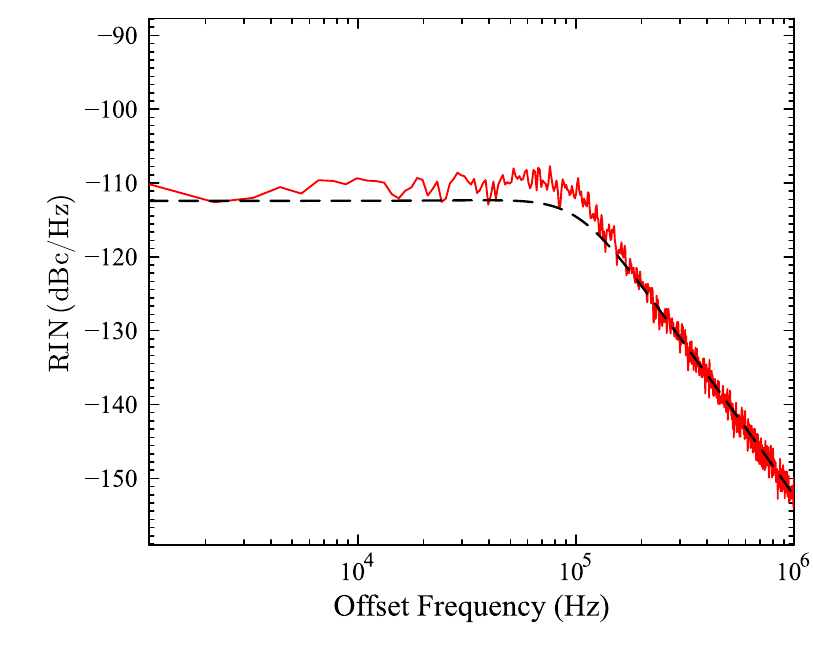}
\caption{Relative intensity noise (RIN) spectrum of the Stokes field.
The red solid line denotes the simulated Stokes RIN, while the black dashed
line indicates the theoretical prediction.
}
\label{RIN}
\end{figure}

To validate the analytical RIN transfer model, we performed numerical simulations by introducing controlled 
amplitude modulation on the input pump field in the discrete-cavity map. This modulation serves as the 
sole external intensity-noise source and drives the coupled optical--acoustic amplitude dynamics. 
From the simulated time-domain trajectories, we extracted the Stokes RIN spectrum by evaluating the power spectral 
density of the normalized Stokes intensity fluctuations.
The result is shown in Fig.~\ref{RIN}, where the red curve denotes the simulated Stokes RIN spectrum and the black 
dashed curve represents the analytical prediction obtained from the linearized transfer function. At low offset 
frequencies, the Stokes RIN remains close to the input pump RIN, indicating that slow pump intensity fluctuations 
are efficiently transferred to the Stokes field. In this regime, the acoustic and optical amplitudes can 
adiabatically follow the pump modulation. 
At higher offset frequencies, however, the Stokes RIN rolls off significantly. This behavior arises from the finite 
response time of the acoustic mode together with the cavity filtering effect, which prevent the Brillouin field 
from tracking rapid pump amplitude fluctuations. The same behavior is also evident from the analytical transfer 
function, whose denominator contains a characteristic pole structure that gives rise to the observed high-frequency 
roll-off. In this respect, the present discrete-cavity result is fully consistent with the behavior previously 
reported for microcavity Brillouin lasers \cite{loh2015noise}.
The good agreement between simulation and theory confirms that the discrete-cavity 
model correctly captures the intensity-noise transfer dynamics.

\section{Conclusion}
In this work, we have developed a discrete-cavity model for free-space
Brillouin lasers. In contrast to conventional continuous-medium
descriptions, the present model treats the short nonlinear
interaction inside the gain medium and the subsequent free-space
propagation as two distinct processes within each cavity round trip.
This separation introduces a temporal asymmetry between the optical and
acoustic dynamics. While the optical fields are stored over the full
round trip, the acoustic field exists only during the finite
interaction interval and continues to decay during the passive
free-space propagation. From the viewpoint of the round-trip optical
dynamics, this effect appears as an enhanced effective damping of the acoustic field.
Based on this discrete-cavity model, we derived analytical results
for the pump-noise suppression ratio, the fundamental intrinsic
linewidth, and the relative intensity noise of free-space Brillouin
lasers. The theoretical predictions are in excellent agreement with
recent experimental observations, indicating that the present model
captures the essential physical mechanism that is absent in
conventional continuous-medium theories. We therefore believe that this
work fills an important gap in the theoretical description of
free-space Brillouin cavities.
Free-space Brillouin lasers are distinguished by their capability for
high-power operation at the watt level. The present results further
show that such systems can simultaneously exhibit strong noise
suppression and ultra-low fundamental noise. This combination of high
power and low noise makes free-space Brillouin lasers highly promising
for practical applications requiring both spectral purity and power
scalability.

\begin{acknowledgments}
The authors acknowledge the supportive research environment provided by the Helmholtz Institute Jena.
\end{acknowledgments}

\appendix
\section{Derivation of the delta-kick interaction update}
\label{app:delta_kick_derivation}
Within the nonlinear gain medium, the slowly varying envelopes of the pump 
field $A_p(t)$, the Stokes field $A_s(t)$, and the acoustic density wave $\rho(t)$ 
obey the driven three-wave coupling equations given in 
Eqs.~(\ref{three_coupled_wave_equations}). Since the interaction takes place only 
over the short time interval $\tau_{\mathrm{int}}$, we approximate the evolution 
during a single pass through the gain medium to first order in 
$\tau_{\mathrm{int}}$. The nonlinear coupling terms are evaluated using the 
pre-interaction amplitudes $A_{p}^{(n)}$, $A_{s}^{(n)}$, and $\rho^{(n)}$, which are assumed 
to remain approximately constant during one pass. This yields
\begin{equation}
\begin{aligned}
A_p^{(n,+)}
&\approx
\left(1-\frac{\gamma}{2}\tau_{\mathrm{int}}\right)A_p^{(n)}
+
\frac{i\omega_{p}\gamma_e}{4n^2\rho_0}\Lambda_{p}\,
\tau_{\mathrm{int}}\,
\rho^{(n)} A_s^{(n)}
\\
&\quad
+
\sqrt{\kappa_{ex}}\,\tau_{\mathrm{int}} S_{\mathrm{in}}
\\
A_s^{(n,+)}
&\approx
\left(1-\frac{\gamma}{2}\tau_{\mathrm{int}}\right)A_s^{(n)}
+
\frac{i\omega_{s}\gamma_e}{4n^2\rho_0}\Lambda_{s}\
\tau_{\mathrm{int}}\,
\rho^{*(n)} A_p^{(n)}
\\
\rho^{(n,+)}
&\approx
\left(1-\frac{\Gamma_B}{2}\tau_{\mathrm{int}}\right)\rho^{(n)}
+
i\frac{\epsilon_0\gamma_e^2q^2}{4\Omega}\Lambda_{\rho}\,
\tau_{\mathrm{int}}\,
A_p^{(n)}A_s^{*(n)}
\end{aligned}
\end{equation}
We then introduce the effective coupling coefficients
\[
g_{p} \equiv \frac{\omega_{p}\gamma_e}{4n^2\rho_0}\Lambda_{p}
\qquad
g_{s} \equiv \frac{\omega_{s}\gamma_e}{4n^2\rho_0}\Lambda_{s}
\qquad
g_\rho \equiv \frac{\epsilon_0\gamma_e^2 q^2}{4\Omega}\Lambda_{\rho}
\]
so that the interaction step can be written in the compact form
\begin{equation}
\begin{aligned}
A_p^{(n,+)}
&=
\left(1-\frac{\gamma}{2}\tau_{\mathrm{int}}\right)A_p^{(n)}
+i g_p\,\tau_{\mathrm{int}}\,\rho^{(n)} A_s^{(n)}
+\sqrt{\kappa_{ex}}\,\tau_{\mathrm{int}}\,S_{\mathrm{in}}
\\
A_s^{(n,+)}
&=
\left(1-\frac{\gamma}{2}\tau_{\mathrm{int}}\right)A_s^{(n)}
+i g_s\,\tau_{\mathrm{int}}\,\rho^{*(n)}A_p^{(n)}
\\
\rho^{(n,+)}
&=
\left(1-\frac{\Gamma_B}{2}\tau_{\mathrm{int}}\right)\rho^{(n)}
+i g_\rho\,\tau_{\mathrm{int}}\,A_p^{(n)}A_s^{*(n)}
\end{aligned}
\end{equation}

In the reduced round-trip description, the interaction segment is treated as a delta-kick, whereas the acoustic decay is taken to accumulate over the full cavity round-trip time $\tau_{\mathrm{rt}}$. Accordingly, the acoustic update after one round trip is written as
\begin{equation}
\rho^{(n+1)}
=
\left(1-\frac{\Gamma_B}{2}\tau_{\mathrm{rt}}\right)\rho^{(n)}
+i g_\rho\,\tau_{\mathrm{int}}\,A_p^{(n)}A_s^{*(n)}.
\end{equation}
The optical fields $A_{p}^{(n,+)}$ and $A_{s}^{(n,+)}$ represent the amplitudes immediately after the localized interaction. During the subsequent free-space propagation over the remainder of the round trip, additional optical attenuation is neglected in the reduced map, so that
\begin{equation}
A_p^{(n+1)} = A_p^{(n,+)} \qquad
A_s^{(n+1)} = A_s^{(n,+)}
\end{equation}
Collecting the optical and acoustic updates, we obtain the reduced round-trip map
\begin{equation}
\begin{aligned}
A_p^{(n+1)}
&=
\left(1-\frac{\gamma}{2}\tau_{\mathrm{int}}\right)A_p^{(n)}
+i g_p\,\tau_{\mathrm{int}}\,\rho^{(n)} A_s^{(n)}
+\sqrt{\kappa_{ex}}\,\tau_{\mathrm{int}}\,S_{\mathrm{in}}
\\
A_s^{(n+1)}
&=
\left(1-\frac{\gamma}{2}\tau_{\mathrm{int}}\right)A_s^{(n)}
+i g_s\,\tau_{\mathrm{int}}\,\rho^{*(n)}A_p^{(n)}
\\
\rho^{(n+1)}
&=
\left(1-\frac{\Gamma_B}{2}\tau_{\mathrm{rt}}\right)\rho^{(n)}
+i g_\rho\,\tau_{\mathrm{int}}\,A_p^{(n)}A_s^{*(n)}
\end{aligned}
\end{equation}
\section{Thermal noise in the discrete cavity map}
\label{app:Langevin_noise}
We start from the deterministic discrete round-trip map
introduced in Sec.~\ref{app:delta_kick_derivation}.
In reality, the acoustic mode is subject to thermal fluctuations.
Because the acoustic field experiences damping at rate $\Gamma_b$,
the fluctuation--dissipation theorem requires the presence of
a Langevin force driving the acoustic degree of freedom.
We therefore modify the acoustic update equation by adding
a stochastic contribution,
\begin{equation}
\rho^{(n+1)}
=
\left(1-\frac{\Gamma_B}{2}\tau_{\mathrm{rt}}\right)\rho^{(n)}
+
i g_\rho\,\tau_{\mathrm{int}}\,A_p^{(n)}A_s^{*(n)}
+
F_\rho^{(n)}.
\label{eq:stochastic_map_app}
\end{equation}
where $F_{\rho}^{(n)}$ denotes the discretized Langevin increment
associated with the $n$-th cavity round trip.
The discrete noise increment is obtained by integrating the
continuous Langevin force over one round-trip interval,
\begin{equation}
\label{Langevin}
F_\rho^{(n)}
=
\int_{n\tau_{\mathrm{rt}}}^{(n+1)\tau_{\mathrm{rt}}}
f_B(t)\, dt,
\end{equation}
where $f_B(t)$ is a complex white-noise force satisfying
\begin{equation}
\langle f_B(t)f_B^*(t')\rangle
=
C_B^{(c)}\,\delta(t-t').
\label{white-noise correlation}
\end{equation}
According to the fluctuation--dissipation relation, the noise
strength of the continuous Langevin force is
\begin{equation}
C_B^{(c)}
=
\frac{k_B T \rho_0 \Gamma_B}{v^2 V_{\mathrm{opt}}},
\label{eq:CB_app}
\end{equation}
where $k_B$ is the Boltzmann constant, $T$ denotes the temperature,
$\rho_0$ is the equilibrium mass density of the medium,
$\Gamma_b$ is the acoustic damping rate,
$v$ is the acoustic velocity, and
$V_{\mathrm{opt}}$ represents the effective optical mode volume
participating in the Brillouin interaction.

Although the Langevin force acts on the acoustic field throughout
the entire round trip, only the fluctuations generated within the
nonlinear interaction segment couple to the optical fields through
the Brillouin process. During the free-space propagation stage the
optical fields are effectively decoupled from the acoustic mode,
so that thermal acoustic fluctuations do not produce additional
optical perturbations. The effective stochastic increment entering
the discrete interaction map can therefore be obtained by integrating
the Langevin force only over the interaction interval
$\tau_{\mathrm{int}}$,
\begin{equation}
F_\rho^{(n)}
=
\int_0^{\tau_{\mathrm{int}}}
e^{-(\Gamma_b/2)(\tau_{\mathrm{int}}-t)}\, f_B(t)\, dt .
\end{equation}
Using the white-noise correlation of $f_B(t)$ in Eq~\ref{white-noise correlation},
the corresponding discrete correlation becomes
\begin{equation}
\langle F_\rho^{(n)}F_\rho^{(m)*}\rangle
=
C_B\,\delta_{nm},
\end{equation}
where the discrete noise strength is
\begin{equation}
C_B
=
\frac{C_B^{(c)}}{\Gamma_B}
\left(1-e^{-\Gamma_B\tau_{\mathrm{int}}}\right)
\simeq C_B^{(c)}\,\tau_{\mathrm{int}} .
\label{eq:CB_discrete}
\end{equation}
This approximation applies when the interaction time is much shorter
than the acoustic damping time, in which case the discrete noise
strength scales linearly with the interaction duration.

\bibliography{ref} 
\end{document}